# Developing clinical informatics to support direct care and population health management: the VIEWER story

Robert Harland[†,1], Tao Wang[†,2], David Codling[1], Catherine Polling[1,2], Matthew Broadbent[1], Holly Newton[1], Yamiko Joseph Msosa[2], Daisy Kornblum[1], Claire Delaney-Pope[1], Barbara Arroyo[1], Stuart MacLellan[1], Zoe Keddie[1], Mary Docherty[1], Angus Roberts[2], Derek Tracy[1,2], Philip McGuire[2,3], Richard Dobson[2,4,5], and Robert Stewart[1,2]

[†] Joint first authors
[1] South London and Maudsley NHS Foundation Trust, UK
[2] Institute of Psychiatry, Psychology and Neuroscience, King's College London, UK
[3] Department of Psychiatry, University of Oxford, UK
[4] Institute of Health Informatics, University College London, UK
[5] Health Data Research UK, UK

*Corresponding authors. Email: Robert.Harland@slam.nhs.uk and tao.wang@kcl.ac.uk

**Abstract**
Electronic health records (EHRs) provide comprehensive patient data which could be better used to enhance informed decision-making, resource allocation, and coordinated care, thereby optimising healthcare delivery. However, in mental healthcare, critical information, such as on risk factors, precipitants, and treatment responses, is often embedded in unstructured text, limiting the ability to automate at scale measures to identify and prioritise local populations and patients, which potentially hinders timely prevention and intervention. We describe the development and proof-of-concept implementation of VIEWER, a clinical informatics platform designed to enhance direct patient care and population health management by improving the accessibility and usability of EHR data. We further outline strategies that were employed in this work to foster informatics innovation through interdisciplinary and cross-organisational collaboration to support integrated, personalised care, and detail how these advancements were piloted and implemented within a large UK mental health National Health Service Foundation Trust to improve patient outcomes at an individual patient, clinician, clinical team, and organisational level.

## Introduction

The National Health Service (NHS) in the United Kingdom (UK) faces increasing pressures[1], with mental health services facing particular challenges due to longstanding resource shortages that hinder their ability to meet the needs of local populations[2–4]. In response, many mental health services have emphasised crisis provision and acute care, which takes a disproportionate amount of a limited budget, while relying on tightly managed access to secondary and inpatient care in a contemporary "productivity agenda"[5,6]. However, this approach fails to address broader systemic challenges in mental healthcare, such as disparities in care quality, inequitable access, high demand for services and the critical need for better integration of mental and physical healthcare, particularly pronounced for individuals with severe mental disorders[7,8].

As part of a broader international move to better meet growing needs in an era of limited resources, there has been a push towards more "integrated care". In the UK Integrated Care Systems (ICSs) have been introduced[4]. These systems prioritise prevention and aim to improve outcomes, reduce health inequalities, and deliver better value for money by mobilising and coordinating the entire healthcare ecosystem, including NHS providers, social care via local councils, voluntary organisations, and other partners[9]. There is robust evidence that this collaborative approach is far more effective in addressing the social determinants of health (SDOH) such as housing, income, employment, and environmental conditions[10]. Services addressing SDOH are essential in improving overall health. Within ICSs, ongoing services for patients discharged from secondary mental health care could include not only services from general practitioners (GPs) but also from the voluntary sector and enhanced primary care.

An essential task for Integrated Care Boards (ICBs) that run ICSs is to use population health management (PHM) approaches to proactively identify individuals at risk and determine the services they may need. Electronic health records (EHRs) capture clinical information at the point of care, providing opportunities to address this task at scale, previously impossible with paper records. However, unlike physical health conditions, where structured data such as blood tests and other metrics provide relatively easily-accessible descriptions of clinical status, mental health assessments rely heavily on patient-reported subjective information. This might include quantitative measures to evaluate mental health presentations, contextual factors, interventions, and outcomes[11]; further, critical information on the clinical presentation, management plan and outcomes is often recorded in free-text formats, such as in clinical case notes and correspondence[12,13], making it challenging to utilise effectively. Thus, a PHM approach focused on stratifying the population according to mental health needs and status might overlook critical factors that influence stratification and intervention strategies until information contained in free-text of EHR is effectively integrated[14].

Recent advances in natural language processing (NLP), a range of techniques that automate the interpreting and understanding of human language, have made it possible to extract medical information buried in clinical text. This has provided enriched clinical datasets, enhancing EHR-based research on mental health conditions[13,15,16]. We present an approach to translate these advances into a platform that helps coordinate and manage mental healthcare. We share our experience of piloting at a large mental health UK NHS Foundation Trust, demonstrating how it can enable effective multi-level resource allocation—from the general population, to specific caseloads, and individual patients. We adopt a component-based approach using open standards to facilitate customisation and replication in other settings. While NHS structures such as ICSs and ICBs shaped the initial development of VIEWER, its agile architecture ensures it can adapt to future changes in the healthcare landscape.

**VIEWER: Leveraging informatics for mental healthcare transformation**

To address the holistic needs of individuals affected by mental health conditions, we developed a digital population health management platform called VIEWER (Visual & Interactive Engagement With Electronic Records) as a proof-of-concept. This platform was developed at the South London and Maudsley NHS Foundation Trust (SLaM), which provides a comprehensive range of NHS mental health services to a local population of 1.3 million residents of four boroughs in South London (Croydon, Lambeth, Lewisham and Southwark), as well as specialist services for children and adults across the UK and beyond.

VIEWER builds upon research-focused informatics resources developed by the NIHR Maudsley Biomedical Research Centre (BRC), which incorporates advanced clinical data analytics tools such as NLP, information retrieval and visual analytics systems[12,17]. The NLP tools enable text-based searches of EHR, identifying meaningful clinical entities beyond the scope of structured data and allowing hitherto infeasible research questions to be addressed, combining both the large sample sizes derived from the EHR with high levels of data granularity (e.g., on symptoms, interventions and other relevant exposures). By integrating and visualizing validated NLP-extracted entities across both physical and mental health domains using CogStack, an open-source, modular clinical information retrieval and visualization system[17], VIEWER curates patient notes and helps clinicians quickly locate relevant information in a timesaving and clinically-actionable format. Technical details and related clinical effectiveness evaluations of VIEWER have been presented in prior work[18]. Building on that foundation, this paper explores how the platform can optimise care delivery across the general population covered by the organisation; specific caseloads, whether delineated by geography, clinical problem-type, or sociodemographic factors;and individual patients. The focus here is to provide insights into enhancing traditional care models through informatics-driven population health management tools and to stimulate further discussion in this domain, rather than presenting empirical findings.

**Multi-level clinical decision support with VIEWER**

Starting in 2019, a project group, including RH and DC (Consultant Psychiatrists), MB (Informatician), and TW and YM (Health Data Scientists) collaborated to develop a three-level proof-of-concept, named VIEWER shown in Figure 1. The platform was designed to facilitate clinical decision making across macro (population), meso (care pathway and caseload) and micro (individual patient) levels. It also supports the targeted outreach from higher clinical tiers, including secondary care and tertiary tier teams, to enhance prevention and provide specialist input, including within primary care (Figure 2).

**Population-level insights:** At the **Population** level, VIEWER visualises the entire living population who have ever utilised SLaM services, which includes approximately 420,000 people. Clinicians responsible, or managers planning services (with an anonymised non-clinical view of patient data), can use the platform to identify areas of varying prevalence and incidence, inequalities and specific diagnostic differences within geographic regions of relevance to their service. An example of this is the visualisation of psychosis incidence in the SLaM South East London catchment in Figure 3(a). Mapping of where new cases are emerging can be combined with other sources of information to inform coordination with both the voluntary and statutory sectors to ensure that services are appropriately located. These services may include social provision around meaningful occupation or employment, cannabis reduction programmes, family support and other pre-clinical social engagement. This approach should incorporate local insights and careful data interpretation, as visualizations may highlight existing differences in service access. Thoughtful attention to underlying inequalities is crucial to addressing disparities in mental health provision rather than inadvertently reinforcing them.

**Pathway-level care management:** At the **Pathway** level, VIEWER visualises the whole population of service users with specific diagnoses, to promote equitable access to care and adherence to relevant standards for all individuals. Among the approximately 20,000 people who have accessed SLaM services with a diagnosis of non-affective psychosis (commonly referred to as schizophrenia-like conditions, with an International Classification of Diseases (ICD)-10 code of F20-29), around two-thirds to three-quarters have been discharged to primary care. However, crucially, with this platform, clinicians are enabled to assess what proportion of this population has received the standard elements of care outlined by National Institute of Health and Care Excellence (NICE) guidelines, including appropriate physical healthcare. It also allows clinicians to identify 1) individuals within specific GP practices who may require additional resources or earlier interventions, such as those discharged on long-term antipsychotic medication, or 2) earlier intervention to prevent later crisis or more coercive care.

**Caseload-level patient management**: Within specific secondary care caseloads, targeting multidisciplinary support to those most in need, while ensuring equity throughout, is an enduring challenge, with traditional models based solely on the limited data of caseload *number*, that does not take morbidity or complexity into account. At the **Caseload** level, VIEWER allows team leaders, consultants and caseworkers to balance their focus more effectively. For example, the platform can track patient presentations to crisis services. In Figure 3(b), a scatter plot illustrates one community team, where each colored dot represents a patient assigned to a specific Care Coordinator. Patients positioned higher on the y-axis (representing crisis service presentations) or the x-axis (indicating bed admissions) signal those who may require more focused attention from the multidisciplinary team (MDT), which includes medical, nursing, psychological, pharmacy, employment, and social support professionals. One goal for the team and the broader system is to work collaboratively to move all patients toward the lower-left corner of the graph, where there is less reliance on crisis care and bed admissions. However, patients with low scores on both axes may still have significant unmet needs, such as social isolation or declining physical health. Thus, a more holistic view of patient care at the Caseload level is essential.

To support holistic care and enhance prevention efforts, the Caseload view also enables real-time tracking of outcomes and evaluation of intervention effectiveness, capturing both quantitative outcomes (e.g., such as physical health metrics including BMI, blood pressure and glucose, and a reduction in admissions) (Figure 3(b)) and qualitative outcomes such as subjective change in well-being and satisfaction (e.g., DIALOG[19], and other Patient-Reported Outcome and Experience Measures or PROMs). This functionality supports coordinated outreach to address areas of low satisfaction, such as challenges with employment or medication, through higher-tier services like employment support services or community pharmacists.

**Individual patient view for enhanced decision-making**: Unlike the interfaces referenced above, which provide population-level health insights by aggregating the latest patient information extracted from unstructured text using NLP and structured data fields, the **Individual** patient view curates concise, meaningful summaries of a patient's entire longitudinal record (Figure 3(c)). This enables clinicians to quickly review essential details at a glance, supporting more informed decision-making and streamlining clinical tasks at the point of care. For example, tasks such as medication reviews, which previously required several hours to complete, can now be performed rapidly through a streamlined dashboard interface. In our preliminary evaluation at prototype stage, the time for each review was reduced from one-two hours to 10-20 minutes, as previously reported[18]. Pharmacists and doctors can now routinely access a comprehensive summary of previous medications, view relevant points of change, and understand the reasons for those changes through visualisations of medication time-series and contextual information from linked source text. Crucially, the platform also overlays service access data with interventions, enabling clinicians to identify which treatments have been most effective for a specific patient or population cohort, and measure any changes effected by the introduction of new or modified interventions. Summaries of past risks and complexities are also available, while physical health results can be viewed with trend lines to show change over time, such as an increase in BMI or a decline in renal function. In high-demand services, this level of curated information can significantly reduce time spent on data retrieval and support more accurate, evidence-based decision-making.

**A use case example of VIEWER in ICS: Individual Placement and Support (IPS)**
It is well-recognised that the digital literacy of many staff in the NHS, and their confidence in working with technology, is often relatively limited[20]. By providing clinicians with interactive access to clinical data about their caseloads, VIEWER facilitated the co-development of practical use cases, enabling a more data-driven approach to daily clinical practice without the need for advanced expertise in data analysis or digital technologies[18]. Specialist outreach teams and activities, including physical health checks, smoking reduction and pharmacy medication reviews, have been developed and funded to support targeted outreach using VIEWER. Over 600 users within SLaM had routinely accessed VIEWER during the pilot. Figure 4 illustrates a use case of VIEWER in IPS, a representative model for ICS.

**Transitioning to routine implementation**
Building on this success, a new Clinical Informatics Service (CIS), clinically led by consultant psychiatrists CP and DC, has received support to migrate VIEWER to new technical infrastructure within the NHS Trust, supported by an operational service. The CIS has three primary aims:

- Release time to care – streamlining clinical data processes to give clinicians more time with patients.
- Address unmet need – helping clinicians to identify opportunities for proactive, targeted outreach and ensure broader access to evidence-based healthcare across the population.

- Motivate action on inequalities – highlighting where inequalities in access, process and outcomes exist and who they exist for, allowing targeted resource allocation to underserved populations.

During the course of the service set-up, progress will be evaluated against these aims based around a theory of change[21]. This includes a series of steps:
1. Building a technical infrastructure robust enough to be relied on for routine clinical use across the Trust;
2. Ensuring clinician and service user confidence in data models and visualisations through regular stakeholder engagement and transparent testing and communication of data definitions;
3. Building a user base beyond relatively data confident early adopters through a user-led redesign of the interface and development of training and support;
4. Iterative co-design of further clinical use cases for dashboards linked to outcomes that benefit patients.

Initial work has established new technical infrastructure and redesigned the VIEWER platform as LUCI (Locating Useful Clinical Information), with deployment to users in community mental health teams currently underway. Building on insights from the piloting of VIEWER, our progress has reinforced that a clinical informatics solution is most effective if it is meaningful to clinical workers, collaboratively developed by a multidisciplinary team, and seamlessly combines the best digital and clinical insights and innovations.

**Discussion**

VIEWER presents an innovative clinical informatics platform and service designed to advance population health management and integrated care in mental health. The platform can support the management of the full continuum of mental health services across different phases of care, including prevention, diagnosis, treatment, follow-up and care coordination. It aims to comprehensively address the health needs of defined populations by enabling prioritization of both prevention and management, integrating social and healthcare services, facilitating innovative care delivery models, and the monitoring of complexity and at-risk patient groups. By facilitating this proactive, population-centred approach, VIEWER supports a move away from the reactive care models traditionally prevalent in mental health[22]. This shift aligns with ICS principles, aiming to deliver care that is more equitable, effective, and sustainable at both individual and population levels.

Unlike other population health management systems that focus narrowly on specific groups (e.g., individuals with severe mental disorders), single-modal data sources (e.g., administrative data), or isolated clinical tasks (e.g., cost auditing)[8,23,24], VIEWER takes a more comprehensive approach to addressing the multidimensional needs of diverse patient groups. It uses clinical data analytics, including NLP and visualisation, to integrate multi-modal data—particularly extensive clinical text—transforming a large-scale routine EHR data into information that is accessible to clinicians in an interactive way, and highlighting actionable insights with the potential to improve patient outcomes. A user-friendly interface facilitates collaborative design with clinicians, enabling the development of clinically meaningful use cases. These innovations mean VIEWER has the potential to extend the scope and effectiveness of population health management within healthcare services, fostering a clinically led, integrated, data-driven approach to improving population health. A further strength of VIEWER is its agility and customisability—underpinned by open-source, lightweight components[18]—which ensures adaptability to future shifts in NHS policy and the wider healthcare environment.

There are challenges to adopting this approach more widely. Developing the VIEWER proof-of-concept relied on research funding and significant input from existing clinical informatics capacity within the Trust's BRC, which provides a level of technical and academic expertise and support not typically available to other healthcare providers. Enabling a move to more proactive and preventative care may be cost-effective in the longer term. However, work towards establishing an operational service and achieving widespread adoption by clinicians requires both substantial additional resourcing to set up, and ongoing commitment to running costs. Any resulting cost savings may not necessarily accrue directly to the secondary mental health care trust implementing these changes. Piloting of VIEWER has highlighted key learning for wider implementation. First, while VIEWER was designed to support a shift in how services operate, success of future implementation will depend on teams being adequately prepared and supported to adapt their workflows. The IPS example illustrates how teams with data-driven workflows have been more readily able to adopt a PHM approach. This does, however, require a wider system commitment to shifting to a more proactive, population health orientated approach reflected in service design, role expectations and job plans. Second, varying levels of digital and data confidence across the workforce impact both adoption and the level of support required. Designing alongside clinicians and piloting with short feedback cycles has been helpful in adapting VIEWER, although involving a greater diversity of clinicians and co-design techniques is needed to improve usability and pilot training and resources. Third, clinical data is inherently complex, and extraction methods, including NLP, have limitations. To ensure safe and effective use, and clinician confidence in tools, clinicians need support in understanding these complexities and attention paid to transparently explaining sources and limitations of data. Finally, the potential for PHM could be greatly enhanced through better data linkage across services within the ICS. Effective mental health tracking requires integrating data from A&E, acute care, primary care, and other sectors to provide a more comprehensive view of patient journeys and service needs.

Beyond these challenges, broader considerations arise when applying this approach in other contexts. Variability in source data fields across Trusts and EHR systems makes standardization difficult, requiring adaptable frameworks and access to expertise about local systems and health records for integration. While the richness of information in clinical free-text may be more consistent across settings, the transferability of NLP models remains an open question, requiring further evaluation. Finally, the needs and clinical focus of teams can differ widely. For example, a Trust like SLaM, which provides services for a population with a high prevalence of psychosis, may have different priorities than others serving different patient populations. Addressing these variations will be crucial for ensuring adaptability and effectiveness across diverse healthcare environments.

**Conclusion**

EHR data offers significant potential to advance population health management and integrated care, particularly in mental health. VIEWER demonstrates how this potential can be realised by presenting EHR information in a way that enables clinicians to effectively manage individual patients while efficiently prioritising needs across their caseloads. Amid strained resources and fragmented care pathways, improving outcomes demands active clinical oversight and proactive management to ensure equitable, evidence-based care for entire populations. VIEWER draws on expertise and resources built up over a long period for research output, enabling engineering with and for reuse of advanced analytics and clinical informatics, which can be employed modularly in different configurations based on the needs of the health service. By presenting information effectively, the platform empowers clinicians to deliver care that is preventative, personalised, and outcome-driven, while facilitating better monitoring of long-term progress.

**Acknowledgement**

The authors would like to express their gratitude to Damian Larkin, Ian Cutting-Jones, Omar Rayner-Andrews, Percy Chinouya, Juliet Hurn, Abigail Bennett, Justine Chow, Man Yuen, Shubhra Mace, Sandra Batchelor, Isaac Enakimio, and many others for their valuable advice and contributions to the project.

**Ethics Statement**

The VIEWER project received approval from multiple information governance bodies within SLaM. The CRIS Oversight Committee approved development of the data model using de-identified data, while the CogStack Oversight Committee approved the use of identifiable data in the VIEWER prototype. Secondary analysis of CRIS data was reviewed and approved by the Oxford Research Ethics Committee C (reference: 18/SC/0372). In addition, a Data Protection Impact Assessment and a Clinical Risk Assessment were completed to ensure compliance with the Trust's data security and patient safety policies during development and deployment.

**Funding Statement**

This study was funded by: (a) Maudsley Charity, (b) the National Institute for Health Research (NIHR) Biomedical Research Centre at South London and Maudsley NHS Foundation Trust and King's College London.
RD is also part-funded by: (c) Health Data Research UK, which is funded by the UK Medical Research Council, Engineering and Physical Sciences Research Council, Economic and Social Research Council, Department of Health and Social Care (England), Chief Scientist Office of the Scottish Government Health and Social Care Directorates, Health and Social Care Research and Development Division (Welsh Government), Public Health Agency (Northern Ireland), British Heart Foundation and Wellcome Trust; (d) The BigData@Heart Consortium, funded by the Innovative Medicines Initiative-2 Joint Undertaking under grant agreement No. 116074. This Joint Undertaking receives support from the European Union's Horizon 2020 research and innovation programme and EFPIA; it is chaired, by DE Grobbee and SD Anker, partnering with 20 academic and industry partners and ESC; and (e) The National Institute for Health Research University College London Hospitals Biomedical Research Centre. These funding bodies had no role in the design of the study, collection and analyses. The views expressed are those of the author(s) and not necessarily those of the NHS, the NIHR or the Department of Health.

RS is part-funded by: i) the National Institute for Health Research (NIHR) Biomedical Research Centre at the South London and Maudsley NHS Foundation Trust and King's College London; ii) an NIHR Senior Investigator Award; iii) the National Institute for Health Research (NIHR) Applied Research Collaboration South London (NIHR ARC South London) at King's College Hospital NHS Foundation Trust; iv) the DATAMIND HDR UK Mental Health Data Hub (MRC grant MR/W014386).

**Competing Interests**

RS declares research support received in the last 36 months from GSK, and royalties from Oxford University Press.

**Data Availability Statement**

All data used in this study are sourced from patient records within SLaM and are not publicly available. Information about the software and tools used in the research is included in the article, with access links provided where available or upon request.

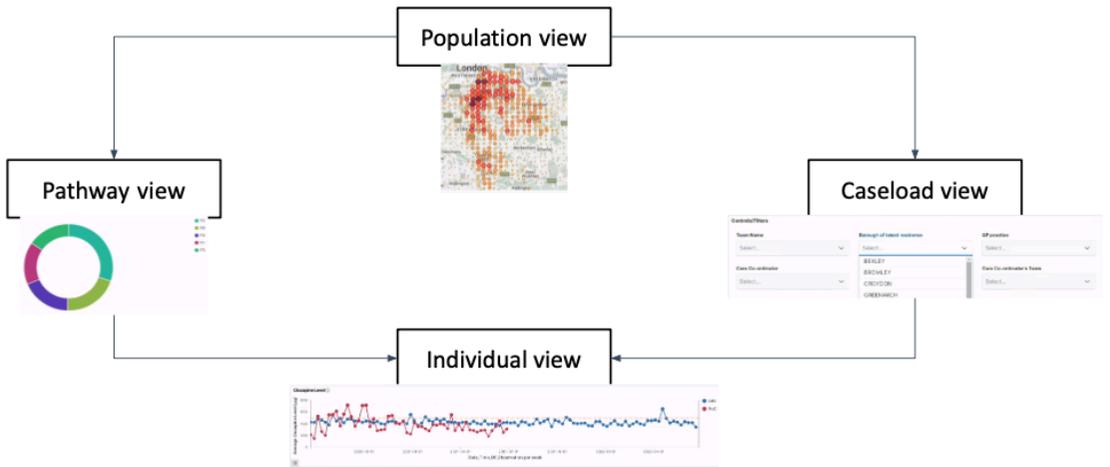

**Figure 1.** Multi-dimensional representation and visualization of EHR data in the VIEWER system, enabling clinicians to easily switch between population, caseload or pathway, and individual patient views.

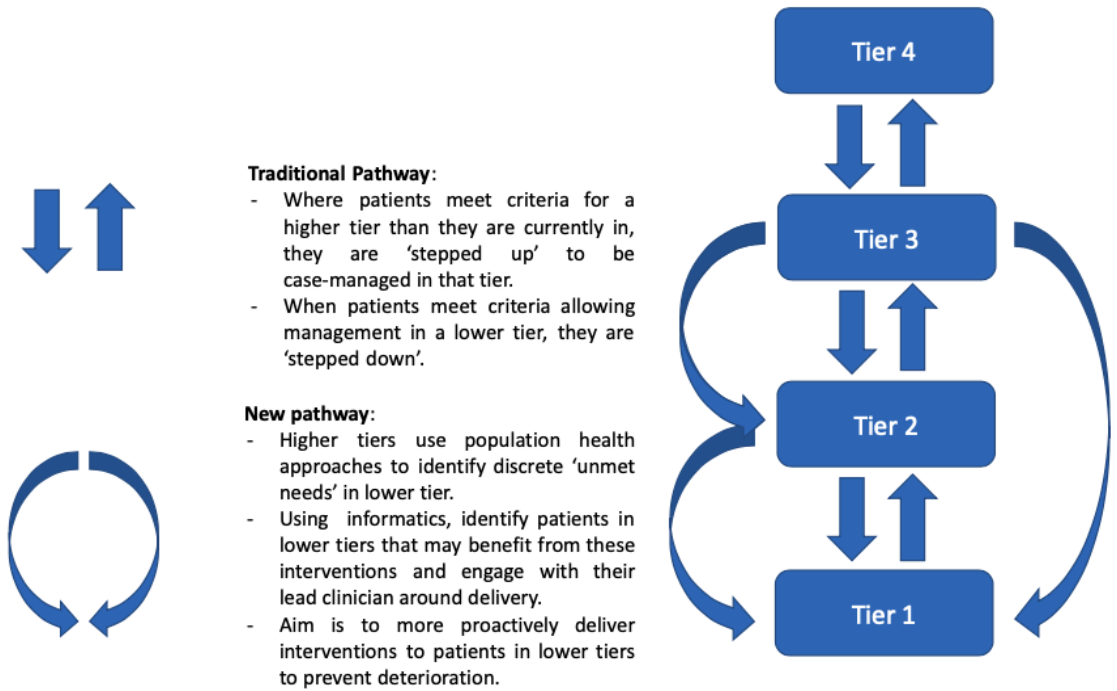

**Figure 2.** Shifting from reactive to proactive care: delivering targeted outreach to populations in need of higher-tier specialist support.

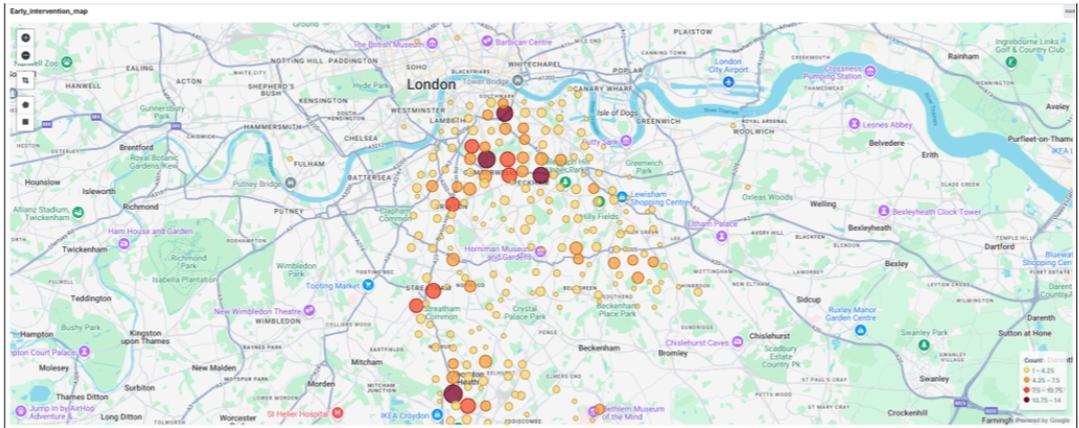

(a) Geographic distribution of populations receiving Early Intervention in Psychosis (EIP) services in the Population View.

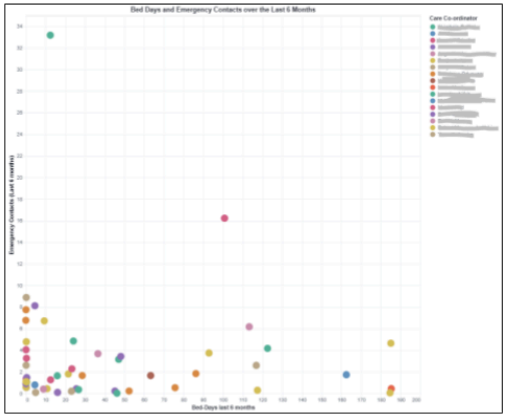

(b) Presentation to crisis services (y-axis) and admission to inpatient beds (x axis) in the Caseload View.

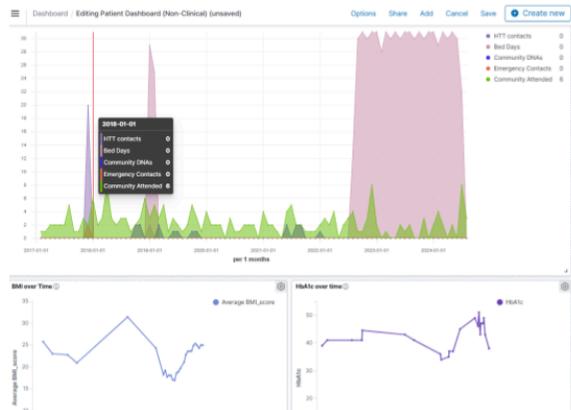

(c) Tracking a patient's antipsychotic use, BMI, and blood glucose levels over time in the Individual Patient View.

**Figure 3.** **(a)** Visualisation of the population currently under SLaM Early Intervention Psychosis (EI) services, allowing a proxy measure of psychosis incidence and targeted integrated prevention. **(b)** A live scatter plot updated daily showing presentation to crisis services (y axis) and admission to beds (x axis). Each dot is one patient and the colours represent care coordinators. Hover over a dot and the name of patient and care coordinator is revealed allowing hyperlink into the Trust's EHR system. This tool can support care coordinators in supervision and better coordination of system resourcing. **(c)** Early version of an individual level visualisation. Medication and other interventions can be overlaid as data over time with use of services, including admissions. Physical health data over time can show changes in BMI, diabetes markers, renal function and medication levels.

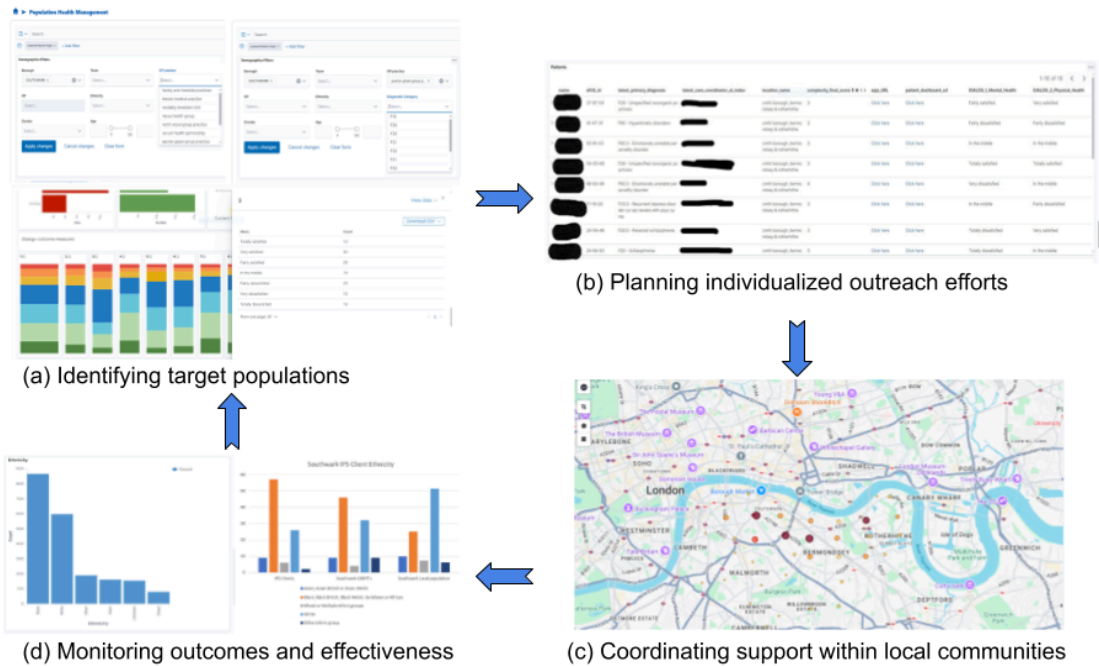

**Figure 4.** IPS workflow enhanced with VIEWER. **(a) Identifying Target Populations**: VIEWER utilizes DIALOG+ health outcome measures, alongside other patient data such as demographics, diagnoses, and clinical team affiliations (including teams in both secondary and primary care), to identify individuals with low job satisfaction or other employment-related needs. **(b) Planning Individualized Outreach**: Employment specialists use VIEWER insights to contact clients directly, offering tailored employment support. For those not ready to engage, clinicians are supported in fostering ongoing discussions about employment goals, enhancing client engagement and promoting holistic, recovery-oriented care. **(c) Coordinating Community-Based Support**: VIEWER's mapping feature highlights areas with high concentrations of IPS clients, enabling employment specialists to prioritize local community spaces and opportunities for sessions. Building connections with these local venues, employers, and areas of opportunity provides clients with non-clinical settings for support, fostering community reintegration and long-term recovery. **(d) Monitoring and Addressing Access Inequalities**: VIEWER tracks disparities in service access across demographic groups, helping target under-served populations in subsequent iterations. This process supports efforts to address racial disparities, share best practices, and integrate race equity into IPS delivery, aligning with SLaM's anti-racism strategy.